\PassOptionsToPackage{table}{xcolor}
\documentclass[sigconf]{acmart}
\AtBeginDocument{%
  }

\setcopyright{acmlicensed}
\copyrightyear{2026}
\acmYear{2026}
\acmDOI{XXXXXXX.XXXXXXX}
\acmConference[Conference acronym 'XX]{Make sure to enter the correct conference title from your rights confirmation email}{June 03--05, 2018}{Woodstock, NY}

\setcopyright{none}                     
\renewcommand\footnotetextcopyrightpermission[1]{} 
\usepackage{pifont}
\newcommand{\cmark}{\ding{51}}%
\newcommand{\xmark}{\ding{55}}%
\usepackage{bbm}
\usepackage[ruled,linesnumbered]{algorithm2e}
\usepackage{multirow}
\usepackage{colortbl}
\usepackage{graphicx}
\usepackage{subcaption} 

\begin{document}

\title{Towards Reliable Negative Sampling for Recommendation with Implicit Feedback via In-Community Popularity}


\author{Chen Chen}
\email{uicetomeetn@gmail.com}
\orcid{0009-0001-6964-9276}
\affiliation{%
  \institution{College of Computer Science and Technology, Jilin University}
  \city{Changchun}
  \state{Jilin}
  \country{China}
}

\author{Haobo Lin}
\email{would1920@gmail.com}
\orcid{0009-0001-0672-0418}
\affiliation{%
  \institution{College of Computer Science and Technology, Jilin University}
  \city{Changchun}
  \state{Jilin}
  \country{China}
}

\author{Yuanbo Xu}
\authornote{Corresponding Author.}
\email{Yuanbox@mails.jlu.edu.cn}
\orcid{0000-0001-8370-5011}
\affiliation{%
  \institution{College of Computer Science and Technology, Jilin University}
  \city{Changchun}
  \state{Jilin}
  \country{China}
}

\renewcommand{\shortauthors}{Chen Chen, Haobo Lin, Yuanbo Xu}

\begin{abstract}
Learning from implicit feedback is a fundamental problem in modern recommender systems, where only positive interactions are observed and explicit negative signals are unavailable. In such settings, negative sampling plays a critical role in model training by constructing negative items that enable effective preference learning and ranking optimization. However, designing reliable negative sampling strategies remains challenging, as they must simultaneously ensure realness, hardness, and interpretability. 
To this end, we propose \textbf{ICPNS (In-Community Popularity Negative Sampling)}, a novel framework that leverages user community structure to identify reliable and informative negative samples. Our approach is grounded in the insight that item exposure is driven by latent user communities. By identifying these communities and utilizing in-community popularity, ICPNS effectively approximates the probability of item exposure. Consequently, items that are popular within a user's community but remain unclicked are identified as more reliable true negatives. Extensive experiments on four benchmark datasets demonstrate that ICPNS yields consistent improvements on graph-based recommenders and competitive performance on MF-based models, outperforming representative negative sampling strategies under a unified evaluation protocol.
\end{abstract}




\maketitle


\section{Introduction}
\begin{figure}[t]
    \centering
    \includegraphics[width=1\linewidth]{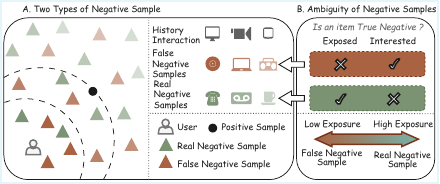}
    \caption {Overview of negative sampling ambiguity and the motivation of ICPNS. (A) Two types of unobserved items under implicit feedback. Triangles with darker colors indicate higher sample hardness; as hardness increases, the likelihood of false negatives also increases. (B) The inherent ambiguity of negative samples: \emph{true negatives} arising from genuine disinterest and \emph{false negatives} caused by lack of exposure. ICPNS approximates item exposure using in-community popularity, treating unclicked but community-popular items as more reliable negatives.}
    \label{fig:overview}
\end{figure}

Implicit feedback such as clicks serves as the foundation for modern Recommender Systems (RS). Unlike explicit feedback such as ratings, implicit feedback suffers from the one-class problem: only positive interactions are observed, while negative signals are absent. Consequently, \textbf{negative sampling} selects unobserved items to represent user disinterest, which is critical for model training (see Figure \ref{fig:overview} A). 
A fundamental challenge in negative sampling lies in the inherent ambiguity of unobserved interactions: an item may be unobserved either because the user genuinely dislikes it or because the user was never exposed to it (see Figure \ref{fig:overview} B). This ambiguity introduces challenges in practice. First, aggressively sampling negatives risks selecting false negatives, exacerbating the False Negative Problem (FNP)\cite{pinsage, fnp1, fnp2}. Second, overly conservative sampling may lead to trivial negatives that slow down optimization. Moreover, when negative samples are selected based on opaque heuristics or model scores alone, it becomes difficult to interpret why a specific item is regarded as a negative instance.
Taken together, these challenges suggest that a \emph{reliable} negative sampling strategy should satisfy several essential requirements. Specifically, it should prioritize samples that are likely to reflect genuine disinterest rather than missing exposure (\emph{realness}), provide sufficiently informative training signals (\emph{hardness}), and offer a transparent rationale for negative selection (\emph{interpretability}). 

Existing negative sampling strategies struggle to satisfy these criteria simultaneously (see Table \ref{tab:negative_sampling_comparison}). The most basic approach, \textbf{Random Negative Sampling (RNS)}\cite{bpr, neumf, lgn, rns3}, randomly selects non-interacted items as negatives. While unbiased and simple to implement, RNS fails all the three criterion due to its randomness. As an improvement, \textbf{Popularity Negative Sampling (PNS)}\cite{pns, soft, fastmf, pns1} assumes that globally popular items are more likely to have been exposed to users. If a user has not interacted with a globally popular item, it is likely a true negative. However, PNS relies on a "global exposure" assumption, ignoring that exposure is highly personalized and community-dependent. It fails to capture the nuance that users are typically exposed to items within their specific interest groups, leading to suboptimal interpretability. \textbf{Hard negative sampling (HNS)}\cite{hns, false, false1, hns1} sample items that the system confuses with positives (e.g., items with higher prediction scores), potentially accelerating convergence since it provide stronger negative signals. However, high-scoring unobserved items are often valid recommendations that the user has not yet seen\cite{pinsage}. Treating them as negatives introduces severe noise, violating realness and degrading (see Figure \ref{fig:overview} D). Some non-sampling work also inspired this work. EXMF\cite{exmf} and FAWMF\cite{fawmf} explicitly model the item exposure, which records the recommended but not interacted items, provides realness of negative samples.

\begin{table}[t]
\centering
\caption{Comparison of representative negative sampling strategies with respect to realness, hardness, and interpretability.}
\label{tab:negative_sampling_comparison}
\small
\begin{tabular}{lccc}
\toprule
\textbf{Strategy} & \textbf{Realness} & \textbf{Hardness} & \textbf{Interpretability} \\
\midrule
RNS   & \xmark & \xmark & \xmark \\
PNS   & \xmark & \xmark & \cmark \\
HNS   & \xmark & \cmark & \xmark \\
ICPNS & \cmark & \cmark & \cmark \\
\bottomrule
\end{tabular}
\end{table}

To bridge these gaps, we revisit a fundamental premise of negative sampling: a non-interaction reflects genuine disinterest only if the user has been exposed to the item. Without exposure, a missing interaction is merely unobserved. From this perspective, identifying reliable negative samples reduces to estimating the probability of item exposure. Motivated by a sociological insight that exposure is structured by user communities\cite{social1, social2}, we propose \textbf{In-Community Popularity Negative Sampling (ICPNS)}. ICPNS models exposure at the community level rather than globally: if an item is popular within a user’s latent community, it is likely to have been exposed to that user. Consequently, an unobserved interaction with a community-popular item constitutes a strong signal of disinterest. Based on this principle, ICPNS samples negatives from items that are popular within a user’s community but individually unobserved, using a lightweight two-stage pipeline that remains computationally efficient.

The main contributions of this work are summarized as follows:
\begin{itemize}
    \item \textbf{A Community-Based View of Exposure.} We reinterpret negative sampling through exposure modeling and argue that exposure is inherently community-structured rather than globally uniform.
    
    \item \textbf{The ICPNS Framework.} We propose In-Community Popularity Negative Sampling, a simple yet effective framework that leverages latent user communities to sample reliable and informative negatives, directly addressing the realness, hardness, and interpretability criteria.
    
    \item \textbf{Empirical Validation.} Extensive experiments on four benchmark datasets demonstrate improvements over representative negative sampling strategies.
\end{itemize}

\begin{figure*}[h]
    \centering
    \includegraphics[width=0.95\linewidth]{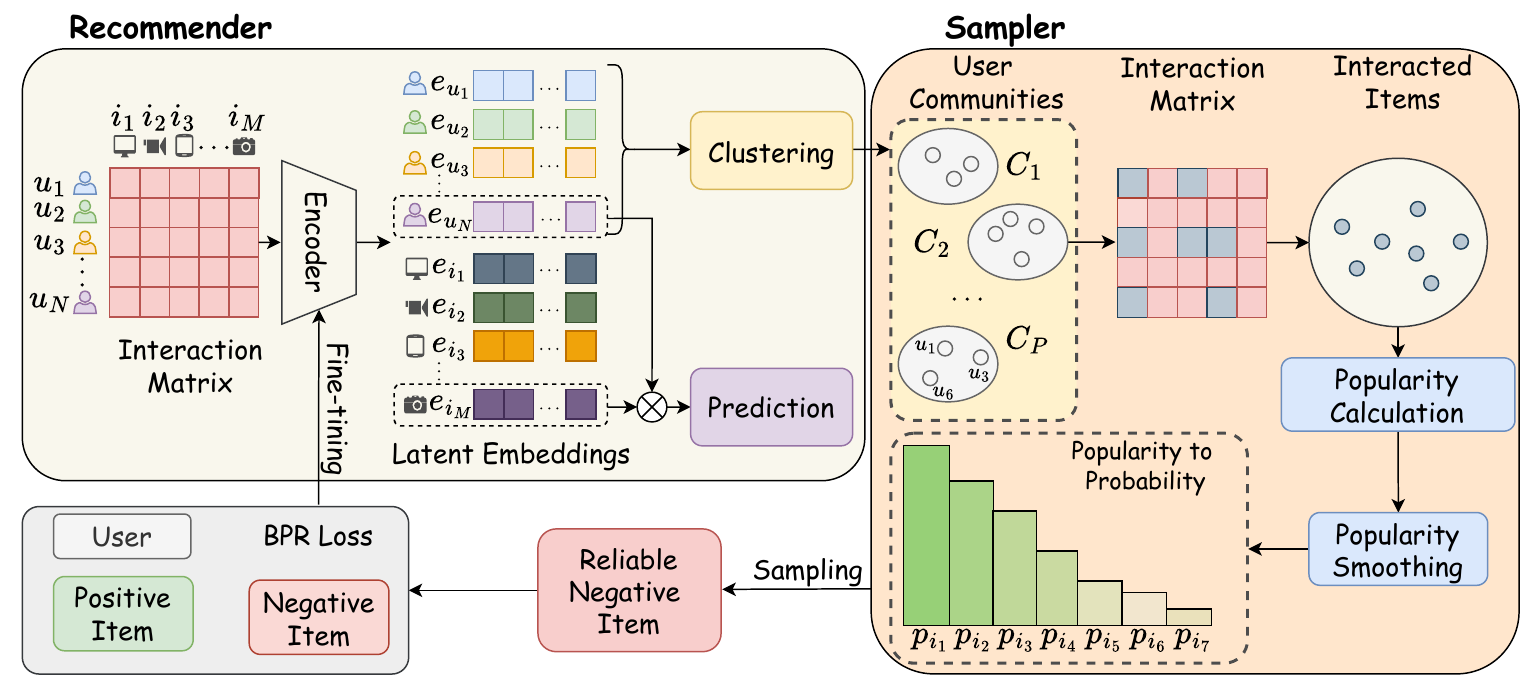}
    \caption{The detailed framework of ICPNS. After encoder is well-pretrained by RNS, the model switches to our ICPNS strategy. Here, users are clustered into latent communities based on pre-trained embeddings, and in-community popularity is leveraged to sample reliable negative items.}
    \label{fig:placeholder}
\end{figure*}

\section{Preliminary}
\subsection{Notations}
Let $\mathcal{U} = \{u_1, u_2, \dots, u_N\}$ and $\mathcal{I} = \{i_1, i_2, \dots, i_M\}$ denote the sets of $N$ users and $M$ items, respectively. The user-item interactions are represented by a binary matrix $\mathbf{X} \in \{0, 1\}^{N \times M}$, where an entry $x_{ui}=1$ indicates an observed interaction (e.g., click, purchase) between user $u \in \mathcal{U}$ and item $i \in \mathcal{I}$, while $x_{ui}=0$ denotes an unobserved interaction. Note that unobserved interactions are inherently ambiguous, representing either a genuine lack of interest (true negative) or a lack of exposure (potential positive). For each user $u$, we denote the set of interacted items as $\mathcal{I}_u^+ = \{i \in \mathcal{I} \mid x_{ui} = 1\}$.

\subsection{Problem Formulation}
The primary objective of recommendation with implicit feedback is to predict a user's preference for items they have not yet interacted with. Formally, the task is to learn a prediction function $f: \mathcal{U} \times \mathcal{I} \to \mathbb{R}$ parameterized by $\mathbf{\Theta}$, which estimates a preference score $\hat{x}_{ui}$ for a user-item pair $(u, i)$:
\begin{equation}
    \hat{x}_{ui} = f(u, i \mid \mathbf{\Theta}).
\end{equation}
For each user $u$, RS aims to generate a personalized ranked list $\mathcal{L}_u$, containing items from the unobserved set $\mathcal{I} \setminus \mathcal{I}_u^+$, sorted in descending order of their predicted scores.

\subsection{Negative Sampling}
During training, a widely adopted objective is the Bayesian Personalized Ranking (BPR) loss\cite{bpr}, which optimizes the pairwise ranking between observed (positive) and unobserved (negative) items. The loss function is defined as:
\begin{equation}
    \mathcal{L}_{\text{BPR}} = \sum_{(u, i, i^-) \in \mathcal{D}_s} - \ln \sigma(\hat{x}_{ui} - \hat{x}_{ui^-}) + \lambda \|\mathbf{\Theta}\|_2^2,
\end{equation}
where $\mathcal{D}_s = \{(u, i, i^-) \mid u \in \mathcal{U}, i \in \mathcal{I}_u^+, i^- \in \mathcal{I} \setminus \mathcal{I}_u^+\}$ represents the set of training triplets. Here, $\sigma(\cdot)$ is the sigmoid function, and $\lambda$ controls the $L_2$ regularization strength.
Negative sampling is the process of selecting a negative item $i^-$ for a given user $u$ and positive item $i$. It can be formally viewed as sampling from a negative distribution $P_{\text{neg}}$:
\begin{equation}
    i^- \sim P_{\text{neg}}(\cdot \mid u, i).
\end{equation}

The sampled item $i^-$ is then used for optimization.

\section{Methodology}

\subsection{Encoder}
ICPNS is a backbone-agnostic negative sampling framework. The role of the encoder is to learn user/item embeddings for (1) constructing a meaningful user embedding space for community identification, and (2) providing preference scores for BPR optimization. Therefore, ICPNS can be instantiated with various backbone such as MF-based and graph-based encoder.

\subsubsection{MF-based Encoder}

Matrix Factorization (MF) models user preferences by assuming that the interaction matrix $\boldsymbol{X} \in \mathbb{R}^{N \times M}$ admits a low-rank approximation. Specifically, MF seeks to decompose $\boldsymbol{X}$ into the product of two low-dimensional embedding matrices:
\begin{equation}
\boldsymbol{X} \approx \boldsymbol{E}_U \boldsymbol{E}_I^\top,
\end{equation}
where $\boldsymbol{E}_U \in \mathbb{R}^{N \times D}$ and $\boldsymbol{E}_I \in \mathbb{R}^{M \times D}$ denote the user and item embedding matrices, respectively, and $D \ll \min(N,M)$ is the embedding dimension.

The predicted preference score between user $u$ and item $i$ is given by the inner product of their corresponding latent embedding:
\begin{equation}
\hat{x}_{ui} = \mathbf{e}_u^\top \mathbf{e}_i,
\end{equation}
where $\mathbf{e}_u$ and $\mathbf{e}_i$ are the $u$-th and $i$-th rows of $\boldsymbol{E}_U$ and $\boldsymbol{E}_I$.

\subsubsection{Graph-based Encoder}

Graph-based encoders learn latent user and item embeddings by performing message passing over the bipartite interaction graph, explicitly modeling high-order collaborative signals beyond direct interactions. Let $D$ denote the embedding dimension. We initialize the embeddings of all users and items in a matrix $\boldsymbol{E}^{(0)} \in \mathbb{R}^{(N+M)\times D}$.

The user--item interaction graph is commonly represented by a bipartite adjacency matrix:
\begin{equation}
    \boldsymbol{A} = \begin{pmatrix}
        \boldsymbol{0} & \boldsymbol{X} \\
        \boldsymbol{X}^\top & \boldsymbol{0}
    \end{pmatrix} \in \mathbb{R}^{(N+M) \times (N+M)},
\end{equation}
where $\boldsymbol{X}$ denotes the interaction matrix. To stabilize message propagation, the adjacency matrix is typically symmetrically normalized as $\tilde{\boldsymbol{A}} = \boldsymbol{D}^{-1/2} \boldsymbol{A} \boldsymbol{D}^{-1/2}$, where $\boldsymbol{D}$ is the diagonal degree matrix with $D_{ii} = \sum_j A_{ij}$.

Graph-based encoders update node embeddings by aggregating embeddings from neighboring nodes according to the graph structure. Without loss of generality, we illustrate this process using LightGCN as an example. 
LightGCN employs a simplified, parameter-free propagation rule given by:
\begin{equation}
    \boldsymbol{E}^{(k+1)} = \tilde{\boldsymbol{A}} \boldsymbol{E}^{(k)}.
\end{equation}

After $K$ propagation layers, embeddings from different layers capture collaborative signals at varying neighborhood ranges. 
A common practice in graph-based recommenders is to aggregate multi-layer embeddings via a weighted sum:
\begin{equation}
    \boldsymbol{E}^{*} = \sum_{k=0}^K \alpha_k \boldsymbol{E}^{(k)},
\end{equation}
where $\alpha_k$ are hyperparameters controlling the contribution of each layer, typically set to $1/(K+1)$. 
The final embedding matrix $\boldsymbol{E}^{*}$ is then split into user embeddings $\boldsymbol{E}^{*}_U \in \mathbb{R}^{N \times D}$ and item embeddings $\boldsymbol{E}^{*}_I \in \mathbb{R}^{M \times D}$, with $\boldsymbol{e}_u$ and $\boldsymbol{e}_i$ denoting the embeddings of user $u$ and item $i$, respectively.

\subsection{ICPNS Negative Sampler}
The proposed ICPNS comprises three key steps: (1) identifying latent user communities, (2) computing item popularity within each community to capture localized exposure patterns, and (3) sampling negative items according to a smoothed in-community popularity distribution.

\subsubsection{Community Identification}
We hypothesize that users located close to each other in the learned embedding space tend to share similar exposure patterns or latent interests. To capture such structure, we apply clustering algorithms to partition the user set $\mathcal{U}$ into $P$ disjoint communities 
$\mathcal{C} = \{C_1, C_2, \dots, C_P\}$, satisfying:
\begin{equation}
    \bigcup_{p=1}^P C_p = \mathcal{U}, 
    \quad 
    C_p \cap C_{p'} = \varnothing \quad \forall p \neq p'.
\end{equation}
The community assignments are fixed after the pre-training stage and remain unchanged during fine-tuning.

\subsubsection{In-community Popularity Computation}
For each identified community $C_p$, we compute an in-community interaction count vector 
$\boldsymbol{v}_p \in \mathbb{R}^M$, where each entry corresponds to the interaction frequency of an item within the community:
\begin{equation}
    (\boldsymbol{v}_p)_j = \sum_{u \in C_p} \mathbbm{1}(x_{uj} = 1),
    \quad \forall j \in \mathcal{I}.
\end{equation}

To mitigate bias toward globally popular items, we further derive a smoothed in-community popularity score.
Specifically, for each community $C_p$, we define a popularity vector $\boldsymbol{s}_p \in \mathbb{R}^M$ as:
\begin{equation}
    (\boldsymbol{s}_p)_j = \big( (\boldsymbol{v}_p)_j \big)^{\alpha},
\end{equation}
where $\alpha \in [0,1]$ is a smoothing hyperparameter. 
When $\alpha < 1$, the influence of items with extremely high interaction counts is suppressed, allowing less frequent but community-relevant items to contribute more effectively to negative sampling.

\subsubsection{Negative Sampling Procedure}
During the fine-tuning stage, ICPNS is employed to sample negative items.
For each positive training pair $(u,i)$ with $u \in C_p$, a negative item $i^-$ is sampled from the set of unobserved items 
$\mathcal{I} \setminus \mathcal{I}_u^+$.
The probability of selecting an item $j$ as a negative sample is defined according to its in-community popularity, normalized over the candidate set:
\begin{align}
    P(i^- = j \mid u \in C_p) 
    &=
    \frac{ (\boldsymbol{s}_p)_j }
    { \sum_{k \in \mathcal{I} \setminus \mathcal{I}_u^+} (\boldsymbol{s}_p)_k },
    \quad
    j \in \mathcal{I} \setminus \mathcal{I}_u^+ .
\end{align}

\subsection{Two-Stage Training Pipeline}

Our model is trained in a two-stage pipeline to maximize the effectiveness of the encoder and the negative sampler.

\textbf{Stage 1: Pre-training.} 
We first pre-train the encoder to learn a stable and informative embedding space.
In this stage, the model is optimized using the BPR loss with RNS, which serves as a standard and robust initialization strategy for collaborative filtering models.

\textbf{Stage 2: Fine-tuning with ICPNS.} 
After pre-training, we fix the community assignments inferred from the pretrained encoder and perform clustering only once considering stability. The model is then fine-tuned by replacing RNS with the proposed ICPNS, while continuing to optimize the BPR loss with negative samples drawn from the in-community popularity distribution.

A detailed algorithmic description of the two-stage training pipeline is provided in Appendix\ref{sec:pipeline}.

\subsection{Complexity Analysis}
We focus on the computational efficiency of negative sampling and exclude the overhead of encoder propagation and clustering during initialization. As shown in Figure\ref{fig:overview} D, all compared methods achieve $O(1)$ time complexity per sampled negative item.

In practice, HNS may incur a larger constant factor due to the need to evaluate preference scores over multiple candidate items. For ICPNS, constant-time sampling is achieved via the Alias Method, which enables $O(1)$ sampling from a discrete non-uniform distribution after preprocessing. Specifically, for each community-level popularity distribution, an alias table is constructed with $O(|\mathcal{I}_c|)$ preprocessing time, where $\mathcal{I}_c$ denotes the item set of community $c$. 

\begin{table}[t]
    \caption{Statistics of four preprocessed datasets.}
    \centering
    \begin{tabular}{c|c|c|c|c}
    \hline
    Dataset & Users & Items & Interactions & Sparsity \\
    \hline
    ML-100K & 940 & 1017 & 80,393 & 91.59\% \\
    ML-1M & 6,034 & 3,124 & 834,449 & 95.57\% \\
    Yelp & 26,420 & 19,921 & 1,285,021 &99.76\% \\
    Beauty & 1,238 & 720 & 25,736 &97.11\% \\
    \hline
    \end{tabular}
    \label{tab:my_label}
\end{table}

\section{Experiment}
In this section, we conduct extensive experiments to systematically evaluate the effectiveness of ICPNS. Specifically, our experimental study is designed to address the following key research questions:
\begin{itemize}
    \item \textbf{Q1:} Can ICPNS consistently improve recommendation performance across diverse datasets and backbone models?
    \item \textbf{Q2:} How does ICPNS compare with existing negative sampling strategies in terms of both effectiveness and computational efficiency?
    \item \textbf{Q3:} Does ICPNS generate negative samples that are both real and hard?
    \item \textbf{Q4:} To what extent are the individual components of ICPNS essential to its overall performance?
    \item \textbf{Q5:} How does different clustering algorithms affect the performance of ICPNS?
    \item \textbf{Q6:} How do key hyperparameters influence the performance of ICPNS?
\end{itemize}

\begin{table*}[t]
    \centering
    \small
    \caption{Performance comparison of various negative sampling strategies across different backbone models. Boldface indicate the best results for each model.}
    \label{tab:sampler_comparison}
    \setlength{\tabcolsep}{2.5pt}
    \begin{tabular*}{\textwidth}{l l @{\extracolsep{\fill}} cccc cccc cccc cccc}
        \toprule
        \multirow{2}{*}{\textbf{Model}} & \multirow{2}{*}{\textbf{Strategy}}
        & \multicolumn{4}{c}{\textbf{ML-100K}}
        & \multicolumn{4}{c}{\textbf{ML-1M}}
        & \multicolumn{4}{c}{\textbf{Yelp}}
        & \multicolumn{4}{c}{\textbf{Beauty}} \\
        \cmidrule(lr){3-6} \cmidrule(lr){7-10} \cmidrule(lr){11-14} \cmidrule(lr){15-18}
        & & Rec & MRR & NDCG & Pre
          & Rec & MRR & NDCG & Pre
          & Rec & MRR & NDCG & Pre
          & Rec & MRR & NDCG & Pre \\
        \toprule

        \multirow{4}{*}{\textbf{BPR}}
        & RNS   & 0.266 & \textbf{0.418} & \textbf{0.264} & \textbf{0.170}
                & 0.172 & 0.416 & 0.238 & \textbf{0.179} 
                & 0.070 & 0.093 & 0.056 & 0.030 
                & 0.231 & 0.145 & 0.146 & 0.043 \\
        & PNS   & 0.253 & 0.408 & 0.254 & 0.162 
                & 0.174 & 0.416 & 0.237 & 0.177 
                & 0.070 & 0.094 & 0.056 & 0.029
                & 0.231 & 0.145 & 0.144 & 0.043 \\
        & HNS   & 0.258 & 0.398 & 0.250 & 0.159 
                & 0.176 & \textbf{0.418} & \textbf{0.239} & 0.178 
                & \textbf{0.085} & \textbf{0.115} & \textbf{0.069} & \textbf{0.036} 
                & \textbf{0.234} & \textbf{0.159} & \textbf{0.155} & \textbf{0.043}\\
        & ICPNS & \textbf{0.269} & 0.413 & 0.257 & 0.162 
                & \textbf{0.176} & 0.415 & 0.238 & 0.178 
                & 0.082 & 0.113 & 0.068 & 0.035
                & 0.224 & 0.147 & 0.147 & 0.041 \\
        \midrule

        \multirow{4}{*}{\textbf{NeuMF}}
        & RNS   & 0.253 & 0.415 & 0.259 & 0.162 
                & 0.161 & 0.380 & 0.217 & 0.166 
                & 0.063 & 0.079 & 0.049 & 0.025 
                & 0.232 & 0.150 & 0.149 & 0.043 \\
        & PNS   & 0.256 & 0.421 & 0.260 & 0.162 
                & 0.160 & 0.377 & 0.214 & 0.164 
                & 0.059 & 0.075 & 0.046 & 0.024 
                & 0.231 & 0.150 & 0.149 & 0.043 \\
        & HNS   & 0.254 & 0.419 & 0.251 & 0.157 
                & 0.183 & 0.434 & 0.245 & 0.179 
                & \textbf{0.069} & \textbf{0.093} & \textbf{0.056} & \textbf{0.029} 
                & 0.229 & \textbf{0.154} & \textbf{0.151} & \textbf{0.044} \\
        & ICPNS & \textbf{0.263} & \textbf{0.424} & \textbf{0.267} & \textbf{0.167} 
                & \textbf{0.188} & \textbf{0.443} & \textbf{0.254} & \textbf{0.186} 
                & 0.064 & 0.079 & 0.049 & 0.025 
                & \textbf{0.234} & 0.144 & 0.147 & 0.043 \\
        \midrule

        \multirow{4}{*}{\textbf{DMF}}
        & RNS   & 0.259 & \textbf{0.414} & \textbf{0.256} & \textbf{0.161} 
                & 0.160 & 0.399 & 0.223 & 0.166 
                & 0.061 & 0.092 & 0.051 & 0.027 
                & 0.232 & \textbf{0.147} & \textbf{0.146} & \textbf{0.044} \\
        & PNS   & 0.256 & 0.405 & 0.250 & 0.158 
                & 0.164 & 0.395 & 0.224 & 0.168 
                & 0.060 & 0.089 & 0.049 & 0.025 & 
                0.198 & 0.127 & 0.125 & 0.036 \\
        & HNS   & 0.248 & 0.400 & 0.246 & 0.154 
                & 0.160 & 0.394 & 0.218 & 0.163 
                & \textbf{0.069} & \textbf{0.094} & \textbf{0.061} & \textbf{0.037} 
                & 0.213 & 0.132 & 0.130 & 0.040 \\
        & ICPNS & \textbf{0.265} & 0.398 & 0.254 
                & 0.158 & \textbf{0.165} & \textbf{0.408} & \textbf{0.228} & \textbf{0.169} 
                & 0.065 & 0.093 & 0.054 & 0.031 
                & \textbf{0.231} & 0.143 & 0.142 & 0.043 \\
        \midrule

        \multirow{4}{*}{\textbf{NGCF}}
        & RNS   & 0.274 & 0.460 & 0.282 & 0.175 
                & 0.175 & 0.420 & 0.239 & 0.178 
                & 0.072 & 0.095 & 0.057 & 0.030 
                & 0.244 & 0.160 & 0.158 & 0.045 \\
        & PNS   & 0.272 & 0.452 & 0.280 & 0.177 
                & 0.176 & 0.412 & 0.236 & 0.177 
                & 0.070 & 0.095 & 0.056 & 0.029 
                & 0.240 & 0.146 & 0.149 & 0.045 \\
        & HNS   & 0.279 & 0.465 & 0.290 & 0.178 
                & 0.192 & 0.450 & 0.261 & 0.192 
                & 0.080 & 0.107 & 0.065 & 0.034 
                & 0.244 & 0.169 & 0.167 & 0.046 \\
        & ICPNS & \textbf{0.323} & \textbf{0.551} & \textbf{0.356} & \textbf{0.219} 
                & \textbf{0.208} & \textbf{0.504} & \textbf{0.304} & \textbf{0.225} 
                & \textbf{0.086} & \textbf{0.123} & \textbf{0.073} & \textbf{0.037} 
                & \textbf{0.262} & \textbf{0.222} & \textbf{0.201} & \textbf{0.050} \\
        \midrule

        \multirow{4}{*}{\textbf{LightGCN}}
        & RNS   & 0.276 & 0.450 & 0.281 & 0.175 
                & 0.183 & 0.429 & 0.249 & 0.186 
                & 0.075 & 0.104 & 0.062 & 0.032 
                & 0.234 & 0.150 & 0.147 & 0.043 \\
        & PNS   & 0.270 & 0.445 & 0.275 & 0.172 
                & 0.182 & 0.431 & 0.248 & 0.184 
                & 0.073 & 0.102 & 0.060 & 0.031 
                & 0.232 & 0.145 & 0.144 & 0.043 \\
        & HNS   & 0.284 & 0.476 & 0.288 & 0.179 
                & 0.206 & 0.478 & 0.279 & 0.202 
                & 0.084 & 0.118 & 0.071 & 0.035 
                & 0.249 & 0.153 & 0.154 & 0.046 \\
        & ICPNS & \textbf{0.311} & \textbf{0.514} & \textbf{0.332} & \textbf{0.204} 
                & \textbf{0.212} & \textbf{0.504} & \textbf{0.304} & \textbf{0.224} 
                & \textbf{0.089} & \textbf{0.126} & \textbf{0.076} & \textbf{0.038} 
                & \textbf{0.260} & \textbf{0.184} & \textbf{0.179} & \textbf{0.050} \\
        \midrule

        \multirow{4}{*}{\textbf{LightGCL}}
        & RNS   & 0.270 & 0.435 & 0.271 & 0.171 & 0.176 & 0.435 & 0.247 & 0.182 & 0.084 & 0.113 & 0.068 & 0.035 & 0.243 & 0.146 & 0.149 & 0.045 \\
        & PNS   & 0.257 & 0.426 & 0.261 & 0.164 & 0.181 & 0.441 & 0.252 & 0.185 & 0.086 & 0.117 & 0.070 & 0.035 & 0.248 & 0.155 & 0.154 & 0.045\\
        & HNS   & 0.272 & 0.445 & 0.269 & 0.166 & 0.1995 & 0.460 & 0.266 & 0.197 & 0.087 & 0.118 & 0.072 & 0.035 & 0.253 & 0.149 & 0.152 & 0.046 \\
        & ICPNS & \textbf{0.316} & \textbf{0.538} & \textbf{0.354} & \textbf{0.227} & \textbf{0.201} & \textbf{0.465} & \textbf{0.271} & \textbf{0.201} & \textbf{0.094} & \textbf{0.137} & \textbf{0.082} & \textbf{0.041} & \textbf{0.261} & \textbf{0.173} & \textbf{0.170} & \textbf{0.049}\\
        \bottomrule
    \end{tabular*}
\end{table*}

\subsection{Experimental Settings}
\subsubsection{Datasets}
We evaluate the proposed ICPNS on four publicly accessible benchmark datasets, covering diverse domains including movies, e-commerce, local businesses and varying data sparsity levels.
\begin{itemize}
\item \textbf{{ML-100K} \& ML-1M}\footnote{https://grouplens.org/datasets/movielens}: Widely used movie recommendation datasets collected by GroupLens Research, containing user ratings on movies.
\item \textbf{Amazon-Beauty}\footnote{https://jmcauley.ucsd.edu/data/amazon/}: A subset of the Amazon product review dataset, representing a sparse e-commerce scenario.
\item \textbf{Yelp}\footnote{https://www.kaggle.com/datasets/yelp-dataset/yelp-dataset/versions/1}: A dataset adopted from the 2018 Yelp Challenge, recording user check-ins at local businesses like restaurants and bars.
\end{itemize}

Preprocessing: Since all four datasets are rating-based, to fit it into implicit feedback task, we first binarized the ratings by excluding interactions below half of the highest rating. We filter out cold-start users and items with less than $k$ interactions ($k=10$ for ML-100K, ML-1M and Beauty, $k=20$ for Yelp). The detailed statistics of the processed datasets are summarized in Table \ref{tab:my_label}.

\subsubsection{Baseline Strategies}
We compare ICPNS against the following negative sampling strategies: 
\begin{itemize}
    \item \textbf{RNS}~\cite{bpr}: Uniformly samples unobserved items as negative instances.
    \item \textbf{PNS}~\cite{pns}: Samples negative items according to their global popularity distribution. Following prior work, we apply popularity smoothing with a smoothing factor of $\alpha=0.1$ to mitigate the dominance of extremely popular items.
    \item \textbf{HNS}~\cite{hns}: Selects hard negatives based on model-predicted preference scores. Following prior work, for each user, we first sample $s=10$ candidate items and then choose the item with the highest predicted score assigned by current model as the negative sample.
\end{itemize}

\subsubsection{Backbone}
To demonstrate the potential of ICPNS on different backbone models, we select models from two mainstream general recommendation categories as follows: 
\begin{itemize}
    \item \textbf{Matrix Factorization (MF-based)}: BPR\cite{bpr} (UAI '09), \\ 
    DMF\cite{dmf} (IJCAL '17), NeuMF\cite{neumf} (WWW '17).
    \item \textbf{Graph-based}: NGCF\cite{ngcf} (SIGIR '19), \\
    LightGCN\cite{lgn} (SIGIR'20), LightGCL\cite{lgl} (ICLR '23).
\end{itemize} 

\subsubsection{Evaluation Protocol}
For dataset partitioning, we employ a randomized splitting strategy, dividing interactions into training, validation, and testing sets with a ratio of 8:1:1, respectively. To evaluate the recommendation performance, we employ four widely adopted metrics: Recall @K (Rec@K), Precision, Normalized Discounted Cumulative Gain @K (NDCG@k), and Mean Reciprocal Rank @K(MRR @K), with $K=10$ by default. For detailed definition, please refer to Appendix \ref{sec:metrics}. Recall and Precision measure the retrieval coverage and accuracy, while NDCG and MRR evaluate the quality of the ranking order. We report the average results over all users in the test set. 

\subsubsection{Implementation Details}
Our framework is implemented using PyTorch and the RecBole\cite{recbole} library. We adopt a two-stage training pipeline. \textbf{Pre-training Stage:} We train the backbone model with RNS for a maximum of 1000 epochs without early stopping to ensure fully converged and stable user embeddings for clustering. \textbf{Fine-tuning Stage:} We switch the sampler to ICPNS and fine-tune the model. The maximum epochs are set to 1000, with an early stopping patience of 20 epochs based on NDCG@10 performance on the validation set.
Hyperparameters: We optimize all models using the Adam optimizer with a learning rate of $1e-3$ and a batch size of 4096. The embedding size is fixed at 64 for all models for fairness. For ICPNS, the number of user communities $P$ is set to 8 for ML-100K/Beauty and 32 for ML-1M/Yelp. The popularity smoothing factor $\alpha$ is set to 0.1. All experiments are conducted on a Linux server equipped with an Intel Xeon Platinum 8352V CPU and an NVIDIA RTX 4090 GPU (24GB).

\subsection{Overall Performance (Q1)}
In this section, we comprehensively evaluate the performance of ICPNS against three baseline negative sampling strategies across six backbone models and four datasets. The detailed performance comparison is presented in Table \ref{tab:sampler_comparison}. 

Overall, ICPNS demonstrates robust improvements over the standard RNS baseline across most settings, validating its effectiveness in mining reliable negatives. Notably, while ICPNS yields competitive results on MF-based models (occasionally underperforming compared to HNS), it consistently and significantly outperforms all baseline strategies on graph-based models. 
This disparity suggests that ICPNS benefits substantially from the structural knowledge encoded by graph neural networks. Since ICPNS relies on clustering learned user embeddings, the high-quality embeddings propagated through the graph structure facilitate more precise clustering, thereby maximizing the efficacy of ICPNS.

\subsection{Training Time Comparison (Q2)} 

\begin{figure}[t]
    \centering
    \begin{subfigure}[t]{0.48\linewidth}
        \centering
        \includegraphics[width=\linewidth]{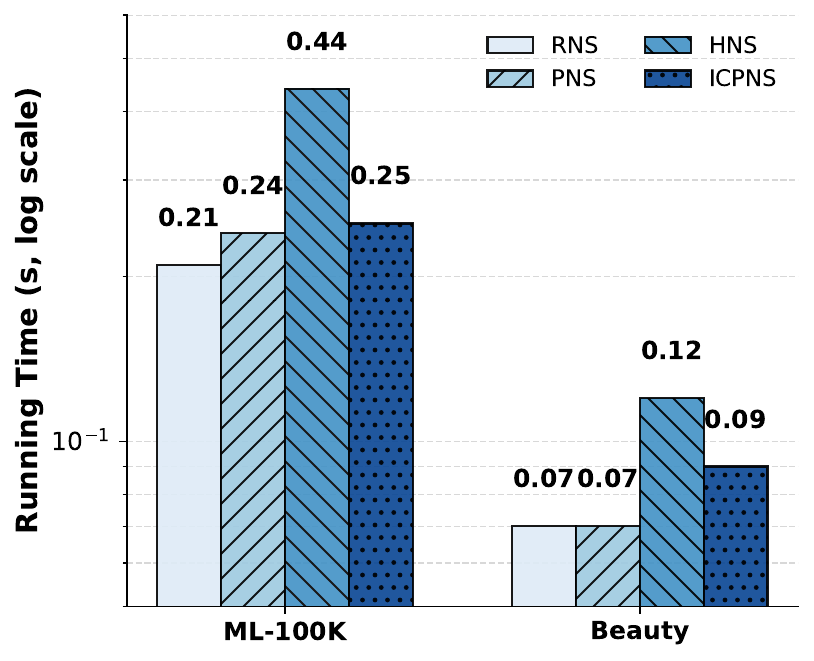}
        \label{fig:rt_small}
    \end{subfigure}
    \hfill
    \begin{subfigure}[t]{0.48\linewidth}
        \centering
        \includegraphics[width=\linewidth]{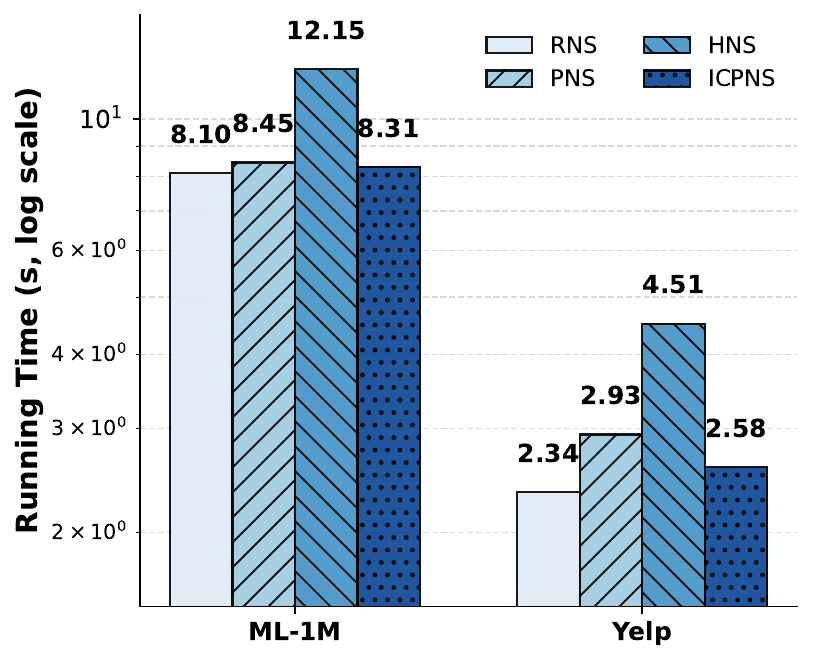}
        \label{fig:rt_large}
    \end{subfigure}

    \caption{Comparison of training time per epoch across datasets. 
    The y-axis is plotted on a logarithmic scale. 
    To improve readability, small and large datasets are reported separately. 
    ICPNS consistently achieves lower computational overhead compared to HNS.}
    \label{fig:rt}
\end{figure}

To evaluate the practical utility of ICPNS, we compare its training efficiency against several representative negative sampling strategies. Figure \ref{fig:rt} illustrates the average training time per epoch on a logarithmic scale. As analyzed in Section 3.4, all considered strategies, including RNS, PNS, HNS, and ICPNS theoretically achieve a sampling complexity of $O(1)$ per negative item. However, the experimental results reveal distinct gaps in actual execution time. Specifically, HNS consistently incurs the highest computational overhead across all datasets. This is attributed to its necessity to compute scores for multiple candidate items to identify hard negatives, which introduces a significantly larger constant factor in practice. 

\subsection{Realness and Hardness of Samples (Q3)}
In this section, we analyze the realness and hardness of negative samples generated by ICPNS in comparison with baseline strategies. 

\begin{figure}[t]  
    \centering
    \begin{subfigure}{0.48\columnwidth}
        \centering
        \includegraphics[width=\linewidth]{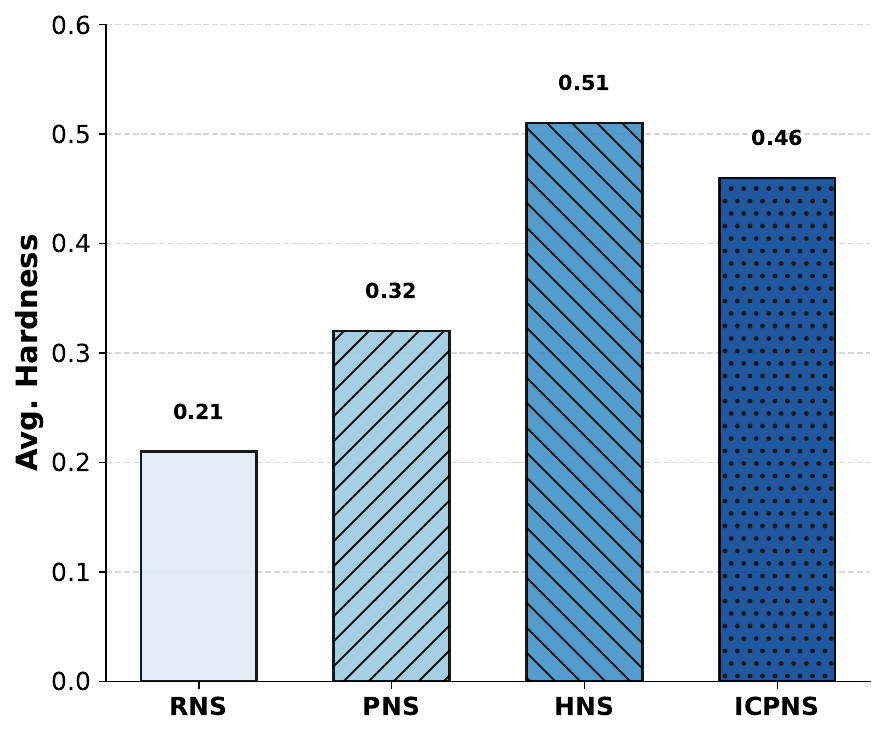}
        \caption{on ML-100K} 
    \end{subfigure}
    \hfill
    \begin{subfigure}{0.48\columnwidth}
        \centering
        \includegraphics[width=\linewidth]{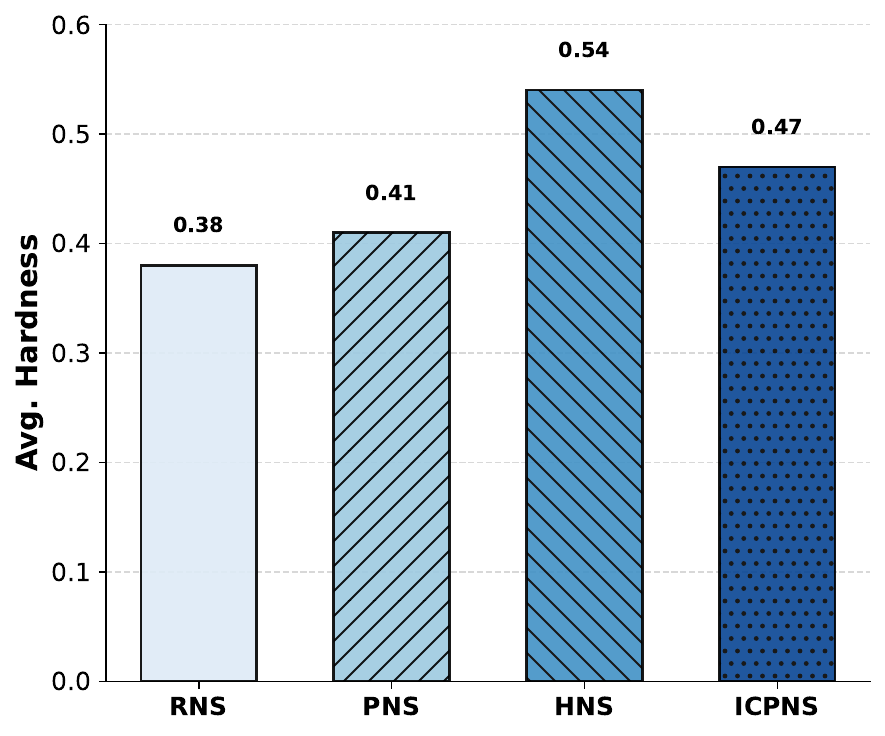}
        \caption{on ML-1M}
    \end{subfigure}
    \hfill
    \begin{subfigure}{0.48\columnwidth}
        \centering
        \includegraphics[width=\linewidth]{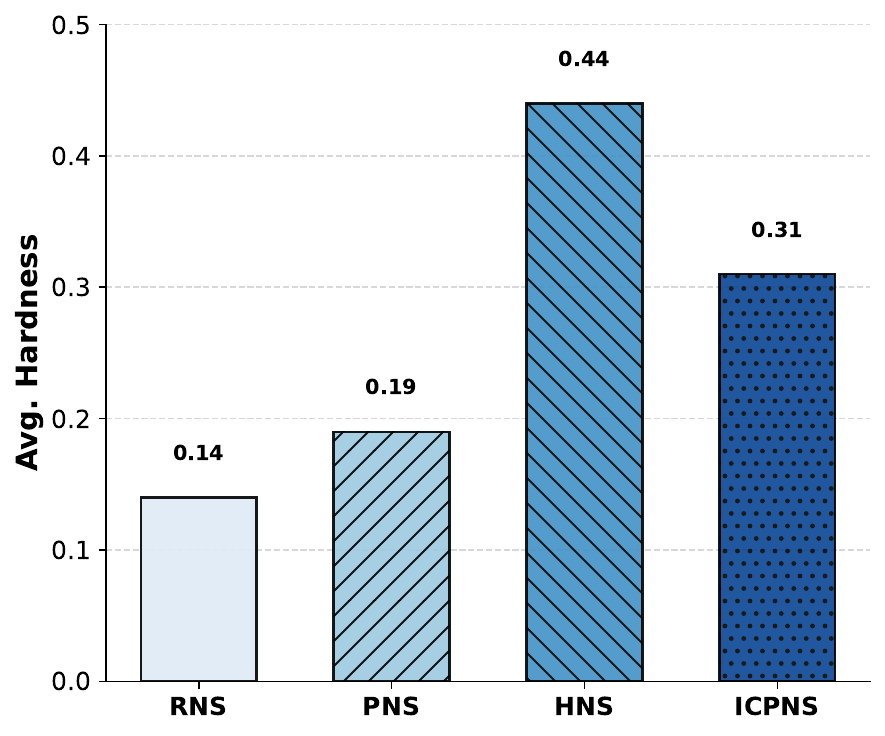}
        \caption{on Yelp}
    \end{subfigure}
    \hfill
    \begin{subfigure}{0.48\columnwidth}
        \centering
        \includegraphics[width=\linewidth]{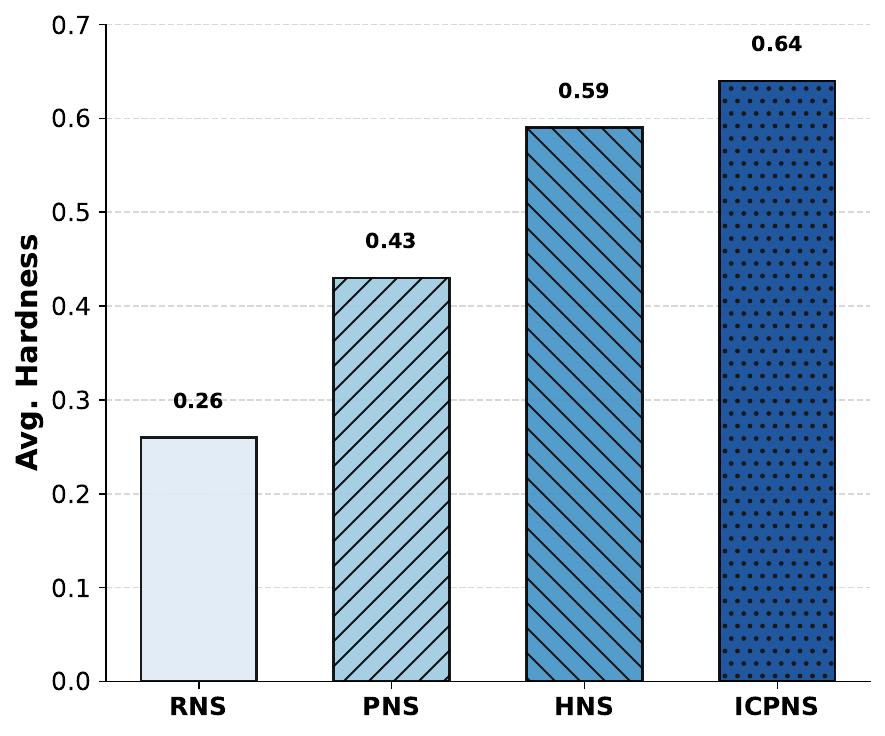}
        \caption{on Beauty}
    \end{subfigure}
    \caption{Comparison of hardness across different negative sampling strategies, averaged over all training epochs across four datasets.}
    \label{fig:hc}
\end{figure}

\textbf{Discussion.} \textbf{Realness:} We introduce an auxiliary metric called \textbf{HoldoutHit@K} to approximate the risk of sampling false negatives. As this metric is not equivalent to exposure-based realness, we report its definition and experimental results in the Appendix \ref{sec:realness}.
\textbf{hardness:} As shown in Figure \ref{fig:hc}, ICPNS consistently generates harder samples compared to RNS and PNS. This superior hardness is an \textbf{inherent benefit of the community-based approach}: items that are popular within a user's specific community usually share high semantic similarity with the user's preferences. Consequently, the model assigns them higher prediction scores, making them naturally hard to discriminate.

\subsection{Ablation Study (Q4)}
To evaluate the effectiveness of each design choice in ICPNS, we perform ablation studies with three variants on the default ICPNS.\footnote{Considering both simplicity and efficiency, we denote the default ICPNS by using LightGCN as encoder and $k$-means as clustering algorithm.}

We consider three ablation variants of ICPNS. 
\begin{itemize}
    \item \textbf{(Variant 1) w/o pretraining}: ICPNS without pretraining the encoder.
    \item \textbf{(Variant 2) w/o clustering}: ICPNS without clustering users into communities. We implement this by setting $P=1$.
    \item \textbf{(Variant 3) w/o smoothing}: ICPNS without popularity smoothing by setting $\alpha=1$.
\end{itemize}

\textbf{Discussion.} The core of ICPNS is clustering users into different communities to leverage in-community popularity. Without pretraining, we cannot obtain high-quality user embeddings, and hence cannot cluster users into the appropriate communities. Without clustering, ICPNS actually degenerates to PNS, which is based on coarse-grained global popularity and fails to learn users' fine-grained preferences. We also notice that ICPNS performs worse without smoothing. Given the inherent noise in implicit feedback, smoothing mitigates popularity bias and introduces necessary randomness. This allows the model to use popularity merely as a soft guidance signal, thereby facilitating the discovery of informative negatives.

\begin{table}[t]
    \centering
    \caption{Ablation study evaluating the impact of key components, including pretraining, user clustering, and popularity smoothing, on recommendation performance across datasets. Default represents standard ICPNS.}
    \label{tab:ab_dataset}
    \setlength{\tabcolsep}{4pt}
    \begin{tabular}{llcccc}
        \toprule
        \textbf{Dataset} & \textbf{Variant} & Rec & MRR & NDCG & Pre \\
        \midrule

        \multirow{4}{*}{\textbf{ML-100K}}
        & Default         & \textbf{0.311} & \textbf{0.514} & \textbf{0.332} & 0.204 \\
        & w/o pretraining & 0.157 & 0.298 & 0.161 & 0.102 \\
        & w/o clustering  & 0.270 & 0.445 & 0.275 & 0.172 \\
        & w/o smoothing   & 0.277 & 0.455 & 0.279 & 0.174 \\
        \midrule

        \multirow{4}{*}{\textbf{ML-1M}}
        & Default         & \textbf{0.212} & \textbf{0.504} & \textbf{0.304} & \textbf{0.224} \\
        & w/o pretraining & 0.166 & 0.422 & 0.237 & 0.176 \\
        & w/o clustering  & 0.182 & 0.431 & 0.248 & 0.184 \\
        & w/o smoothing   & 0.183 & 0.513 & 0.287 & 0.202 \\
        \midrule

        \multirow{3}{*}{\textbf{Yelp}}
        & Default         & \textbf{0.089} & \textbf{0.126} & \textbf{0.076} & \textbf{0.038} \\
        & w/o pretraining & 0.051 & 0.070 & 0.041 & 0.021 \\
        & w/o clustering  & 0.073 & 0.102 & 0.060 & 0.031 \\
        & w/o smoothing  & 0.079 & 0.113 & 0.065 & 0.033 \\
        \midrule

        \multirow{3}{*}{\textbf{Beauty}}
        & Default         & \textbf{0.260} & \textbf{0.184} & \textbf{0.179} & \textbf{0.050} \\
        & w/o pretraining & 0.185 & 0.107 & 0.110 & 0.032 \\
        & w/o clustering  & 0.232 & 0.145 & 0.144 & 0.043 \\
        & w/o smoothing   & 0.251 & 0.176 & 0.171 & 0.048 \\
        \bottomrule
    \end{tabular}
\end{table}

\subsection{Impact from the Quality of User Communities (Q5)}
\begin{table}[t]
    \centering
    \caption{Performance comparison of ICPNS under different clustering algorithms, reflecting the impact of clustering quality.}
    \label{tab:clustering}
    \setlength{\tabcolsep}{4pt}
    \begin{tabular}{llcccc}
        \toprule
        \textbf{Dataset} & \textbf{Clustering} & Rec & MRR & NDCG & Pre \\
        \midrule

        \multirow{5}{*}{\textbf{ML-100K}}
        & $k$-means & 0.311 & 0.514 & 0.332 & 0.204 \\
        & GMM       & 0.335 & 0.560 & 0.374 & 0.240 \\
        & DBSCAN    & 0.330 & 0.555 & 0.371 & 0.239 \\
        & AC        & 0.347 & \textbf{0.603} & \textbf{0.404} & \textbf{0.251} \\
        & SC        & \textbf{0.350} & 0.596 & 0.401 & 0.247 \\
        \midrule

        \multirow{5}{*}{\textbf{Beauty}}
        & $k$-means & 0.260 & 0.184 & 0.179 & 0.050 \\
        & GMM       & 0.266 & 0.226 & 0.206 & 0.051 \\
        & DBSCAN    & \textbf{0.294} & \textbf{0.257} & 0.231 & \textbf{0.059} \\
        & AC        & 0.277 & 0.270 & \textbf{0.234} & 0.056 \\
        & SC        & 0.280 & 0.225 & 0.209 & 0.054 \\
        \bottomrule
    \end{tabular}
\end{table}

In this section, we detailedly evaluate how the clustering algorithms effects the performance of ICPNS. We evaluate the performance of ICPNS using LightGCN as the encoder on ML-100K and Beauty with different clustering algorithms. \textbf{Partition-based}: $k$-means. $k$-means partitions users into $k$ disjoint clusters by minimizing the within-cluster variance. \textbf{Probability-based}: Gaussian Mixed Model (GMM). GMM models the user embedding distribution as a mixture of Gaussian components, allowing soft cluster assignments. \textbf{Density-based}: DBSCAN. DBSCAN identifies user communities as high-density regions in the embedding space and naturally treats sparse users as noise. \textbf{Hierarchical}: Agglomerative Clustering (AC). Agglomerative clustering builds a hierarchy of user communities by iteratively merging the most similar clusters based on a linkage criterion. \textbf{Graph-based}: Spectral Clustering (SC). Spectral clustering operates on a similarity graph constructed from user embeddings and partitions users by leveraging the eigenstructure of the graph Laplacian.

\textbf{Discussion.} The performance comparison is shown in Table \ref{tab:clustering}. Notably, with advanced clustering algorithm being adopted, the performance of ICPNS is accordingly upgrading. This indicates the learned community structure is vital in ICPNS since we need to calculate the in-community popularity.

\subsection{Hyperparameter Analysis (Q6)}
In this section, we demonstrate the impact of two key hyperparameters in ICPNS on ML-100K and Beauty. Due to space constraint, result on Yelp and ML-1M is shown in Appendix \ref{sec:appendix_hyper}. 
\begin{figure}[t]  
    \centering
    
    \begin{subfigure}{0.48\columnwidth}
        \centering
        \includegraphics[width=\linewidth]{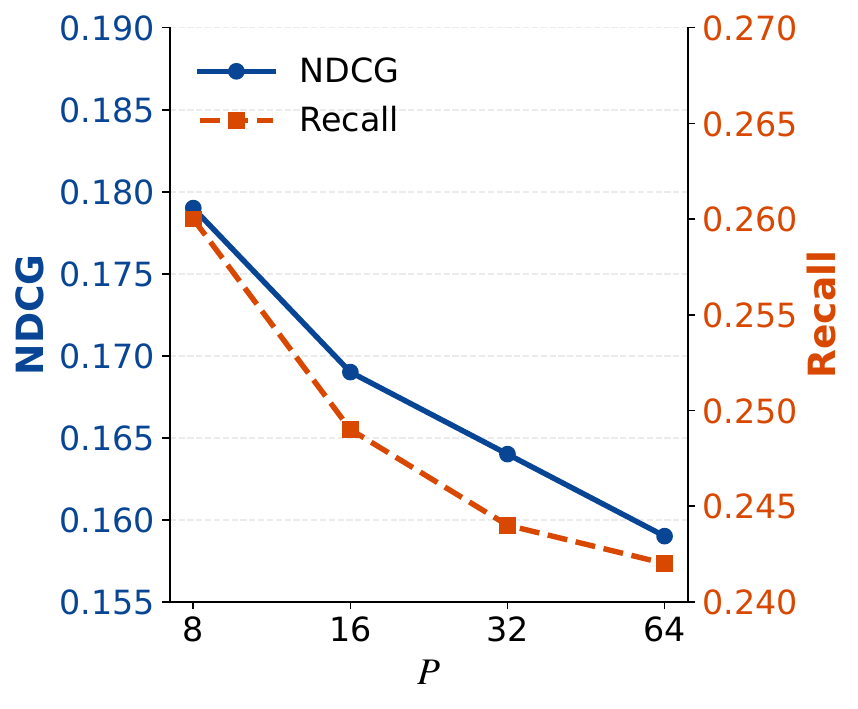}
        \caption{on Beauty}
    \end{subfigure}
    \hfill
    \begin{subfigure}{0.48\columnwidth}
        \centering
        \includegraphics[width=\linewidth]{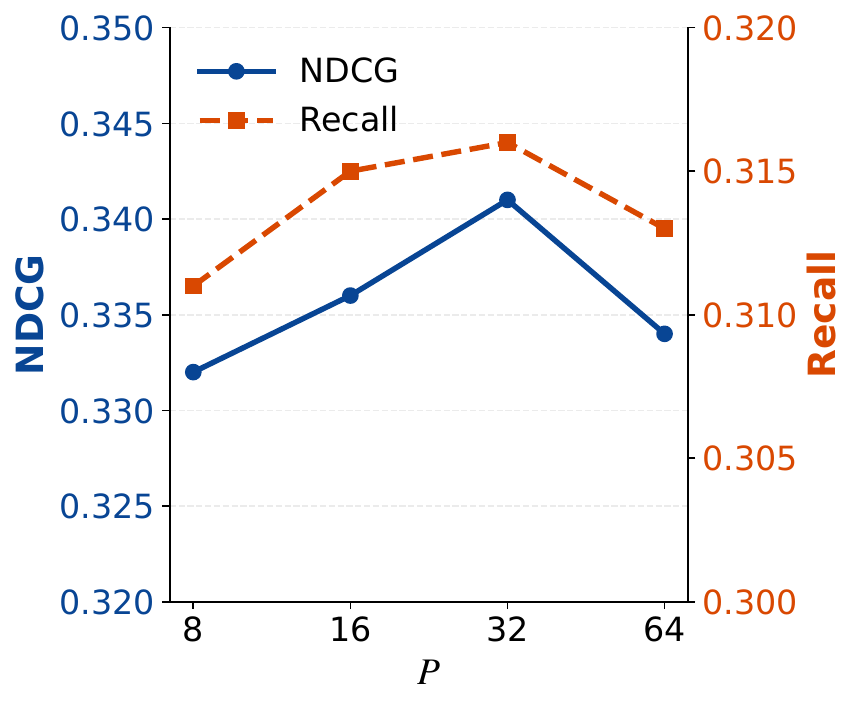}
        \caption{on ML-100K}
    \end{subfigure}

    \vspace{2pt} 

    \begin{subfigure}{0.48\columnwidth}
        \centering
        \includegraphics[width=\linewidth]{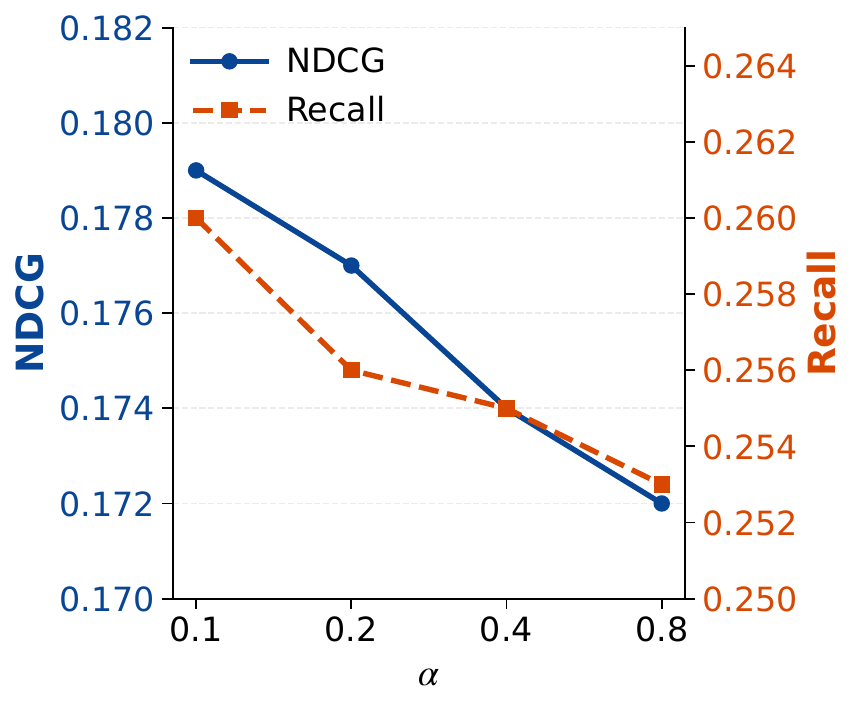}
        \caption{on Beauty}
    \end{subfigure}
    \hfill
    \begin{subfigure}{0.48\columnwidth}
        \centering
        \includegraphics[width=\linewidth]{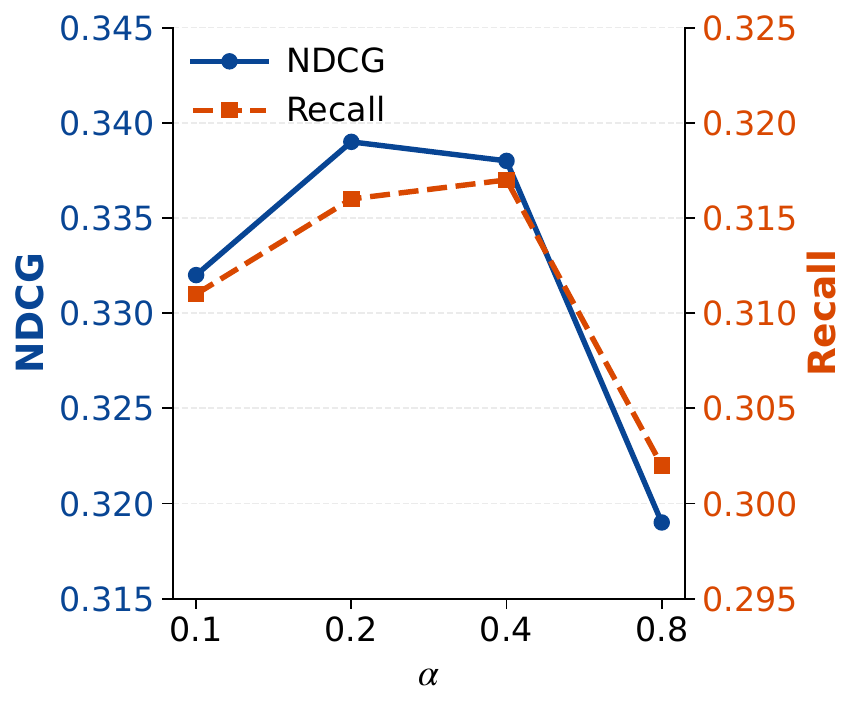}
        \caption{on ML-100K}
    \end{subfigure}

    \caption{Parameter sensitivity analysis of ICPNS on Beauty and ML-100K. The top row (a–b) illustrates the impact of $P$ , while the bottom row (c–d) displays the effect of $\alpha$.}
    \label{fig:hyper_comp}
\end{figure}
\subsubsection{Impact of the Number of Communities \textit{P}}
We investigate the effect of the number of user communities by varying $P \in \{8, 16, 32, 64\}$ and. As shown in Figure \ref{fig:hyper_comp} (a) and (b), the optimal community granularity is highly dataset-dependent. On the relatively small Beauty dataset, performance reaches its peak at $P=8$ and consistently degrades as $P$ increases, indicating that excessive partitioning leads to overly sparse communities and unstable signals for negative sampling. In contrast, for the larger ML-100K dataset, performance attains its maximum at $P=32$, suggesting that finer-grained community structures are beneficial for capturing more nuanced user preference patterns when sufficient data are available.

\subsubsection{Impact of the Popularity Smoothing Factor \boldmath{$\alpha$}}
We further examine the sensitivity of the popularity smoothing factor $\alpha$, which controls the degree of smoothing applied to in-community popularity. Figure \ref{fig:hyper_comp} (c) and (d), reports NDCG and Recall under $\alpha \in \{0.1, 0.2, 0.4, 0.8\}$. Across both datasets, the model achieves its best performance with relatively small values of $\alpha$ from 0.1 to 0.4. As $\alpha$ increases to 0.8, performance consistently declines, indicating that excessive smoothing weakens informative popularity signals. These results highlight the necessity of moderate popularity smoothing for effectively mitigating popularity bias.

\subsection{Case Study}

\begin{figure}[t]  
    \centering
    
    \begin{subfigure}{0.45\columnwidth}
        \centering
        \includegraphics[width=\linewidth]{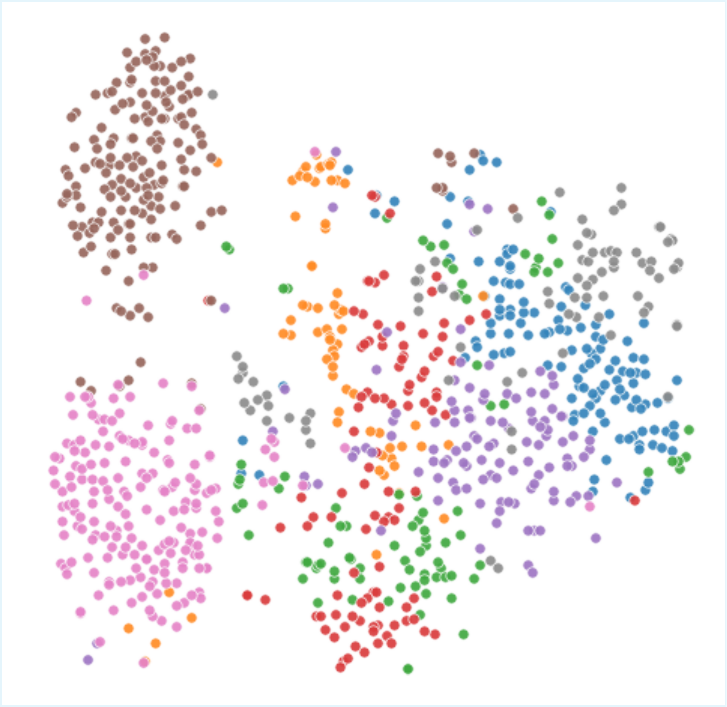}
        \caption{Before Fine-tuning}
    \end{subfigure}
    \hfill
    \begin{subfigure}{0.45\columnwidth}
        \centering
        \includegraphics[width=\linewidth]{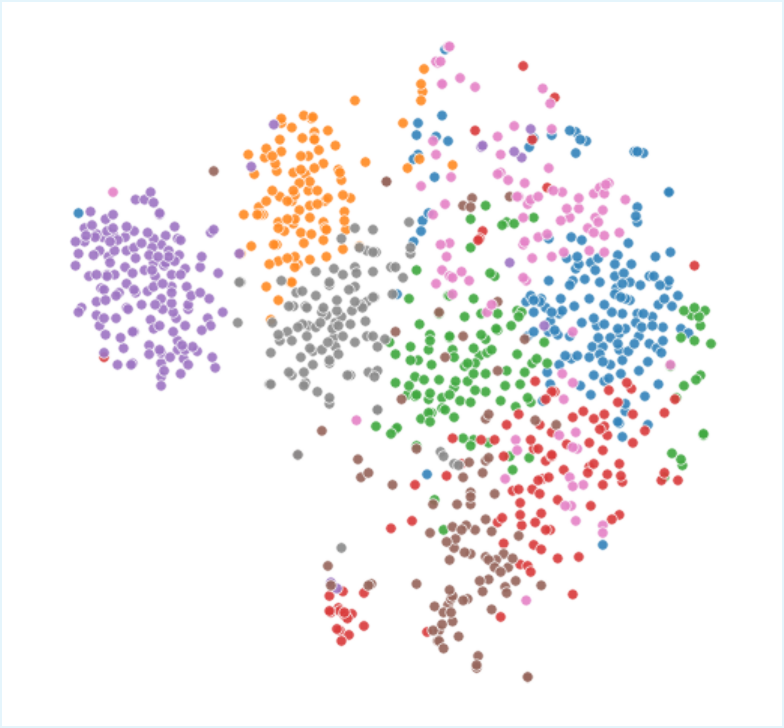}
        \caption{After Fine-tuning}
    \end{subfigure}
    \caption{Visualization of community structure before and after fine-tuning with ICPNS at $P=8$.}
    \label{fig:cluster_comp}
\end{figure}

\begin{table}[t]
\centering
\small
\setlength{\tabcolsep}{6pt}
\renewcommand{\arraystretch}{1.15}
\caption{Clustering quality metrics before and after fine-tuning on ML-100K.
Lower Silhouette and Calinski–Harabasz scores indicate a more dispersed embedding structure. $\Delta$ denotes \textit{After} minus \textit{Before}.}
\label{tab:cluster_quality_before_after}
\begin{tabular}{ccc ccc}
\toprule
\multicolumn{3}{c}{\textbf{Silhouette}} 
& \multicolumn{3}{c}{\textbf{Calinski--Harabasz}} \\
\cmidrule(lr){1-3}\cmidrule(lr){4-6}
\textbf{Before} & \textbf{After} & $\boldsymbol{\Delta}$ 
& \textbf{Before} & \textbf{After} & $\boldsymbol{\Delta}$ \\
\midrule
0.132 & 0.091 & -0.041 & 83.231 & 65.624 & -18.393 \\
\bottomrule
\end{tabular}
\end{table}

\begin{table}[t]
\centering
\small
\caption{Community 3 and its top-5 popular movies.}
\begin{tabular}{c|c}
\toprule
\multicolumn{2}{c}{\textbf{Community 3}} \\
\midrule
\textbf{Movie Name} & \textbf{Genres} \\
\midrule
\textit{The Net} & \cellcolor{gray!20} Sci-Fi, Thriller \\
\textit{Hoodlum} & \cellcolor{gray!20} Crime, Film-Noir \\
\textit{Hellraiser} & \cellcolor{gray!20} Horror, Sci-Fi \\
\textit{Aladdin and the King of Thieves} & \cellcolor{green!20} Children's \\
\textit{Bound} & Crime, \cellcolor{gray!20} Thriller \\
\bottomrule
\end{tabular}
\label{tab:community_example}
\end{table}

To further demonstrate the effectiveness of ICPNS, we provide a case study on the ML-100K dataset with $P=8$.

Table~\ref{tab:cluster_quality_before_after} and Figure~\ref{fig:cluster_comp} illustrate the embedding structure before and after ICPNS fine-tuning.
After applying ICPNS, the embedding space becomes more dispersed, with reduced cluster compactness, indicating that the model shifts from coarse community-level grouping toward more fine-grained and personalized user representations.

As shown in Table~\ref{tab:community_example}, \textit{Aladdin and the King of Thieves}, a \colorbox{green!20}{children-}\\\colorbox{green!20}{oriented} movie, appears in Community~3 alongside predominantly \colorbox{gray!20}{adult-oriented} films such as \textit{Hellraiser} and \textit{Hoodlum}. Despite its high popularity, \textit{Aladdin and the King of Thieves} is unlikely to match the preferences of users favoring adult-oriented content within this community and therefore constitutes an informative negative candidate. By accounting for in-community popularity, ICPNS is able to identify such preference-misaligned popular items as effective negative samples.

\section{Related Work}
Learning from implicit feedback is challenging due to the absence of explicit negative signals. To address this issue, existing negative sampling strategies can be broadly categorized into three groups: RNS, PNS and HNS. Below, we briefly review these strategies and representative recommender systems that adopt them.

\textbf{Random Negative Sampling.} RNS uniformly samples negative instances from items with no observed user interactions\cite{bpr, neumf, lgn, rns3, xu1, xu2}. Owing to its simplicity and low computational overhead, RNS has been widely adopted in collaborative filtering and graph-based recommendation models, especially in large-scale settings. Random sampling also introduces stochasticity, which can enhance item exploration and alleviate overfitting. However, RNS does not guarantee the quality or informativeness of sampled negatives, as many randomly selected items may be trivially irrelevant to the user. Several studies such as TDM\cite{tdm} attempt to improve uniform random sampling under structured settings, such as hierarchical item trees or sequential recommendation.

\textbf{Popularity Negative Sampling.} PNS selects negative samples based on the popularity of items, that is, the more popular the item is, the more likely it is to be selected as the negative samples. Relying on the assumption that the popularity of items may demonstrate users’ global preferences, a series of popularity-based methods such as WBPR\cite{wbpr}, HRNN\cite{hrnn} and BINN\cite{pns1} typically assign sampling weights to items based on their frequency. This encourages the recommender to incorporate more popular items as negative samples into the training process. For quick updating the recommender parameters with the new incoming data, eALS\cite{elas} assigns the weight of missing data based on the popularity of items by an incremental update strategy. SoftRec\cite{softrec} allows some unobserved items to have non-zero supervisory signals by generating soft targets with popularity distribution, thereby incorporating more unobserved knowledge into the training process. 

\textbf{Hard Negative Sampling.} HNS takes another philosophy, it hypothesizes that any unobserved item should not be ranked higher than any observed positive item and subsequently oversamples items top-ranked by the recommender from the randomly selected candidates. To be more specific, HNS generally selects the item from the randomly selected item candidates who has the highest rating given by current system\cite{hns, ahns, fashion, pinsage, mcrec, survey1}. Notably, PinSage\cite{pinsage} argues that the highest-scored item is likely to be a potentially positive sample, leading to the false negative problem.

\section{Conclusion}
In this work, we revisit negative sampling from the perspective of reliability and formalize three essential criteria (realness, hardness, and interpretability) that a negative sampling strategy should satisfy. Motivated by the observation that item exposure is structured by latent user communities, we propose ICPNS, a simple yet effective framework that leverages in-community popularity signals to identify reliable negative samples. Extensive experiments show that ICPNS yields consistent gains on graph-based models and competitive performance on MF-based models.
By explicitly connecting negative sampling with exposure modeling at the community level, this work offers a new and interpretable direction for designing reliable negative sampling strategies in recommender systems.

\bibliographystyle{ACM-Reference-Format}
\bibliography{sample-base}

\appendix

\section{Two Stage Training Pipeline}
\label{sec:pipeline}

\begin{algorithm}[ht]
\small
\caption{Two-Stage Training Pipeline with ICPNS}
\label{alg:icpns_pipeline}
\KwIn{
Interaction matrix $\mathbf{X}$,
number of communities $P$,
smoothing factor $\alpha$,
learning rate $\eta$
}
\KwOut{Optimized model parameters $\boldsymbol{\Theta}^*$}
\BlankLine

Initialize model parameters $\boldsymbol{\Theta}$\;
\While{not converged}{
    Sample a mini-batch of positive interactions 
    $\mathcal{B} = \{(u,i)\mid x_{ui}=1\}$\;
    For each $(u,i)\in\mathcal{B}$, sample $i^- \sim \text{Uniform}(\mathcal{I}\setminus\mathcal{I}_u^+)$\;
    Update $\boldsymbol{\Theta}$ by minimizing the BPR loss\;
}

\BlankLine
Extract user embeddings $\mathbf{E}_U \leftarrow \text{Encoder}(\mathbf{X}; \boldsymbol{\Theta})$\;
Cluster users into $P$ disjoint communities $\{C_1,\dots,C_P\}$ using $\mathbf{E}_U$\;
\For{$p=1$ \KwTo $P$}{
    Compute in-community interaction counts 
    $\boldsymbol{v}_p$ via Eq.~(10)\;
    Compute smoothed popularity 
    $\boldsymbol{s}_p$ via Eq.~(11)\;
    Construct alias table $\mathcal{T}_p$ from $\boldsymbol{s}_p$\;
}

\BlankLine
\While{not converged}{
    Sample a mini-batch of positive interactions 
    $\mathcal{B} = \{(u,i)\mid x_{ui}=1\}$\;
    \ForEach{$(u,i)\in\mathcal{B}$}{
        Identify community $C_p$ such that $u\in C_p$\;
        Sample $i^- \sim \text{Alias}(\mathcal{T}_p)$ 
        \textbf{subject to} $i^- \in \mathcal{I}\setminus\mathcal{I}_u^+$\;
    }
    Update $\boldsymbol{\Theta}$ by minimizing the BPR loss\;
}
\Return{$\boldsymbol{\Theta}^*$}
\end{algorithm}

\section{Additional Experiment}

\subsection{Realness of Samples}
\label{sec:realness}
In implicit feedback recommendation, negative samples are not directly observable, which makes it challenging to assess their reliability. Beyond evaluating negative sampling strategies through ranking performance, we introduce an auxiliary diagnostic metric to examine whether sampled negatives correspond to withheld positive interactions.

We define \textbf{HoldoutHit} to quantify the probability that a sampled negative item appears in the holdout positive set. For each user, we record the items sampled as negatives during training and check whether these items occur among the user’s validation interactions. Since the validation set consists of positive interactions withheld from training, an overlap indicates that the sampled negative item is a potential false negative with respect to future interactions.

Formally, let $\mathcal{N}_u^{(K)}$ denote the set of $K$ negative items sampled for user $u$ during training, and let $\mathcal{S}_u^{\mathrm{val}}$ denote the set of validation positives for the same user. The HoldoutHit metric is defined as
\begin{equation}
\mathrm{HoldoutHit@}K
= \mathbb{E}_{u} \left[
\frac{|\mathcal{N}_u^{(K)} \cap \mathcal{S}_u^{\mathrm{val}}|}{K}
\right].
\end{equation}

\begin{figure}[t]
    \centering
    \includegraphics[width=0.9\linewidth]{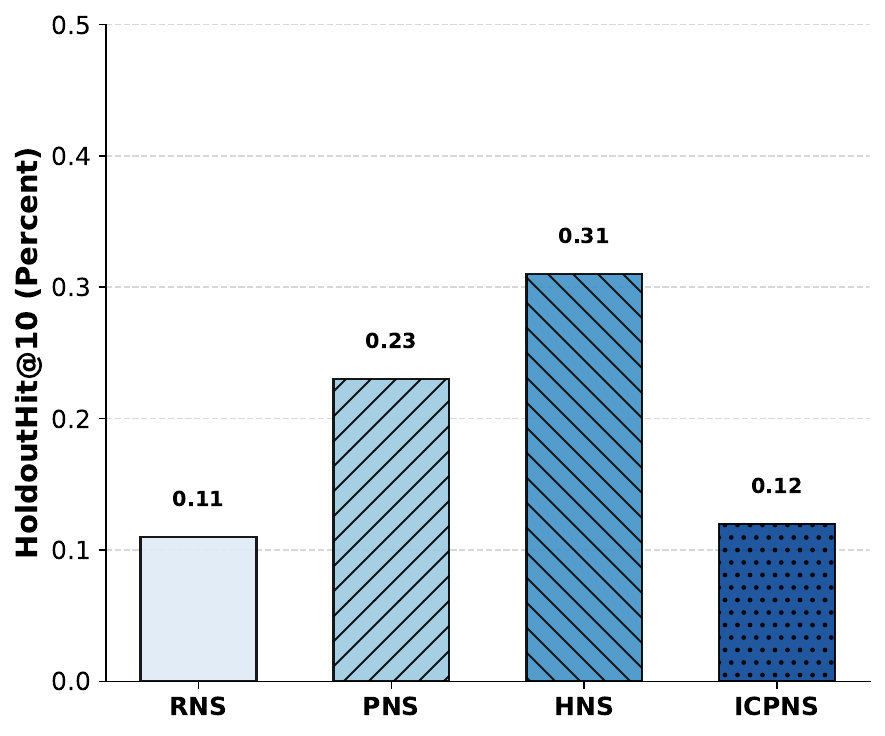}
    \caption{on ML-100K}
    \label{fig:hoh}
\end{figure}

The HoldoutHit@10 results are reported in Figure \cite{hoh}.HNS attains the highest value, indicating a higher likelihood of sampling items that later appear as validation positives. PNS also exhibits an elevated HoldoutHit, reflecting the tendency of popular items to reoccur in future interactions. In contrast, ICPNS yields a lower HoldoutHit than both HNS and PNS, while remaining higher than random sampling.

Random negative sampling (RNS) achieves the lowest HoldoutHit score. This behavior can be attributed to its uniform sampling strategy and the high sparsity of the dataset, under which randomly sampled items are unlikely to coincide with withheld positives. However, a low HoldoutHit does not necessarily imply higher negative-sample realness, as it may also result from sampling items that are irrelevant or unexposed to the user.

Overall, HoldoutHit captures the risk of sampling potential false negatives with respect to future interactions, but it is not equivalent to the exposure-based notion of realness considered in this work and should be interpreted as a complementary diagnostic metric.

\subsection{Hyperparameter Analysis}
\label{sec:appendix_hyper}
\begin{figure}[t]  
    \centering
    
    \begin{subfigure}{0.48\columnwidth}
        \centering
        \includegraphics[width=\linewidth]{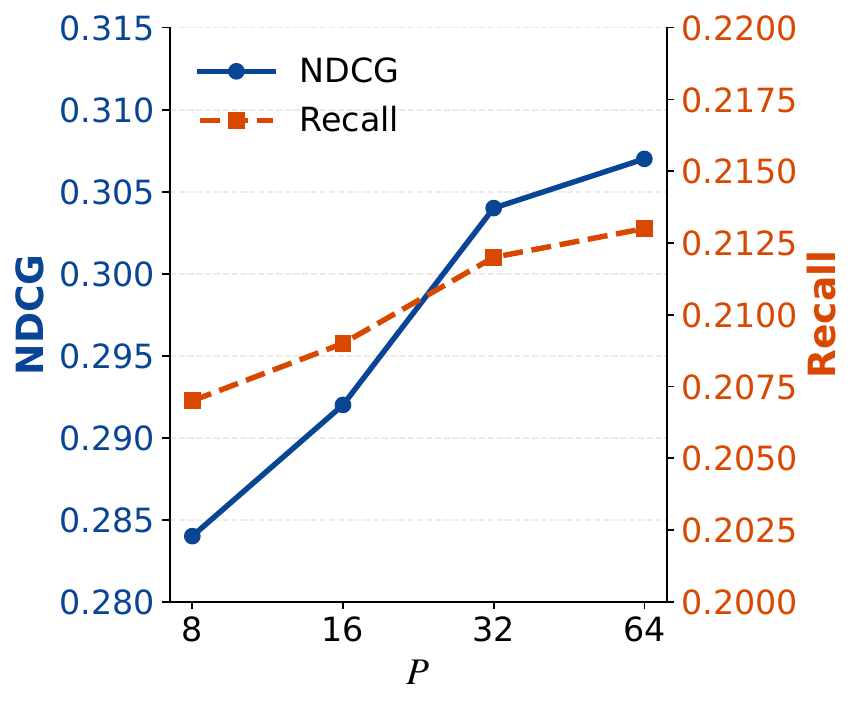}
        \caption{on ML-1M}
    \end{subfigure}
    \hfill
    \begin{subfigure}{0.48\columnwidth}
        \centering
        \includegraphics[width=\linewidth]{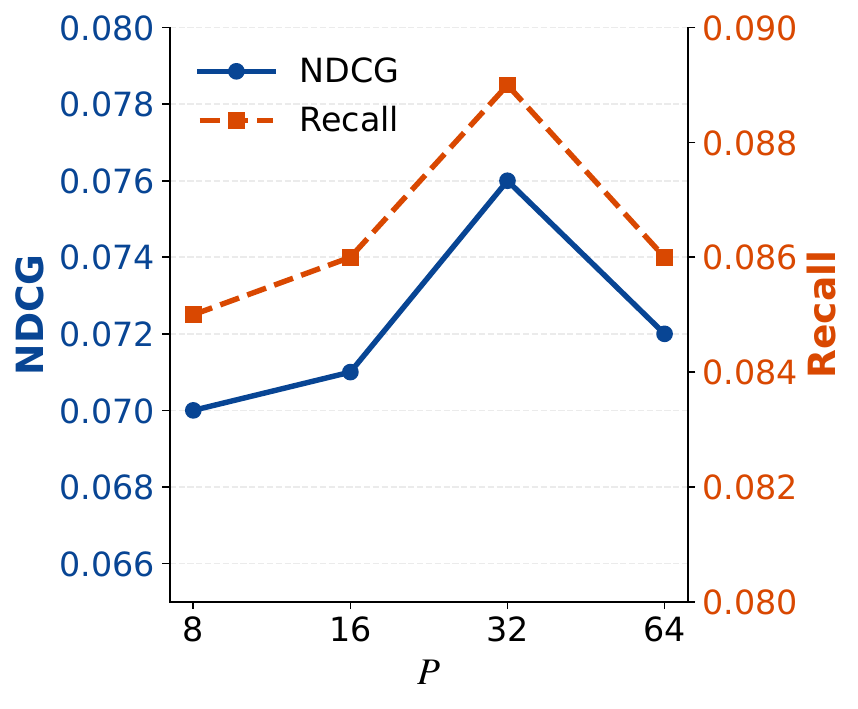}
        \caption{on Yelp}
    \end{subfigure}

    \vspace{2pt} 

    \begin{subfigure}{0.48\columnwidth}
        \centering
        \includegraphics[width=\linewidth]{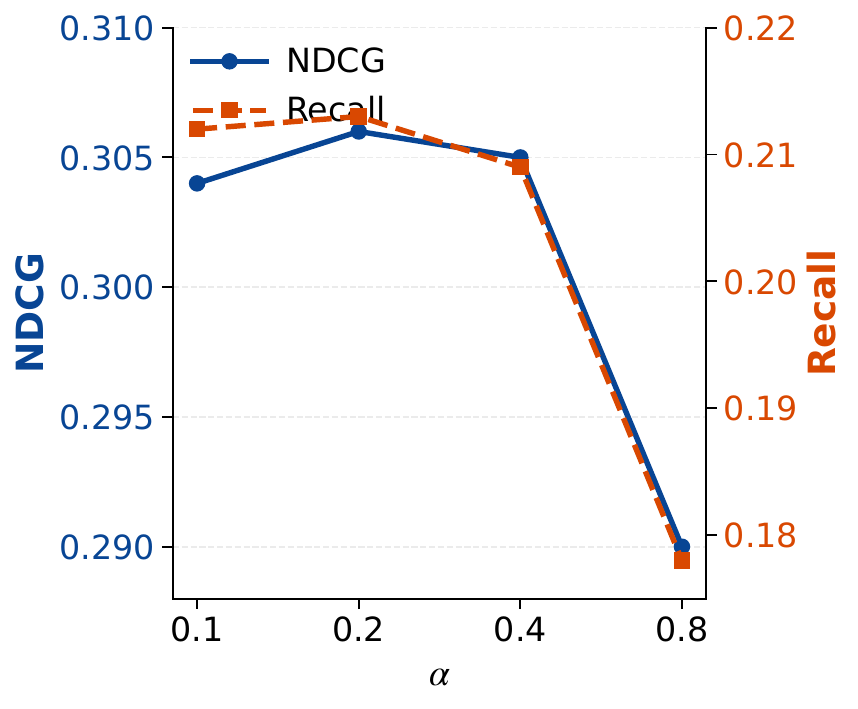}
        \caption{on ML-1M}
    \end{subfigure}
    \hfill
    \begin{subfigure}{0.48\columnwidth}
        \centering
        \includegraphics[width=\linewidth]{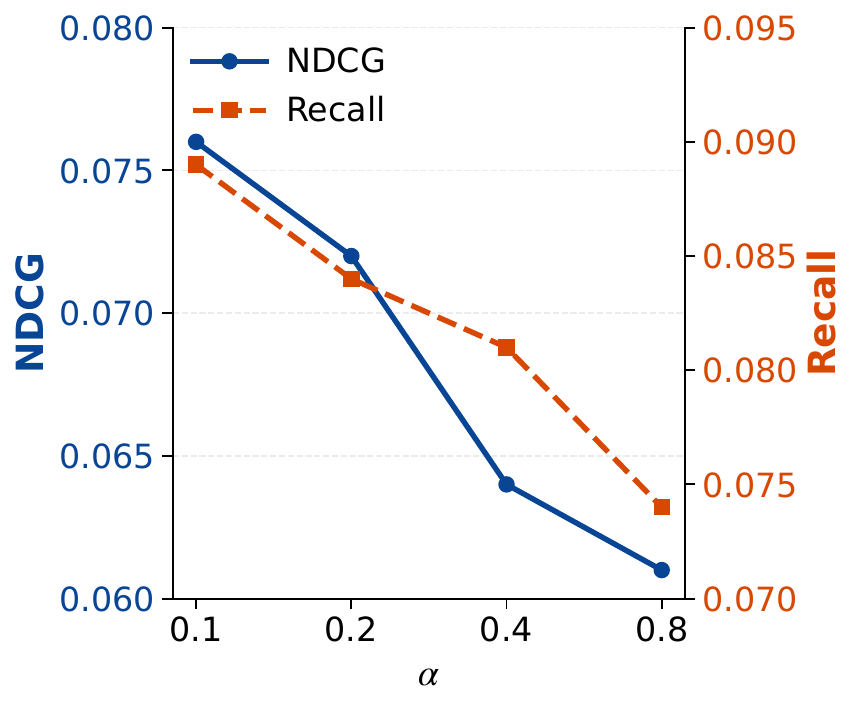}
        \caption{on Yelp}
    \end{subfigure}

    \caption{Parameter sensitivity analysis of ICPNS on ML-1M and Yelp. The top row (a–b) illustrates the impact of $P$ , while the bottom row (c–d) displays the effect of $\alpha$.}
    \label{fig:hyper_comp_big}
\end{figure}

\textbf{Number of Communities \(P\).} 
As illustrated in Figure~\ref{fig:hyper_comp_big}(a) and (b), performance on both datasets improves as \(P\) increases from small to moderate values and then saturates or slightly declines. On ML-1M, both NDCG and Recall consistently increase with larger \(P\), indicating that finer-grained community structures are effective when sufficient interaction data are available. On Yelp, performance peaks at \(P=32\) and exhibits a mild degradation at \(P=64\), suggesting diminishing returns from overly fine partitioning. Overall, moderate-to-large values of \(P\) provide stable and competitive performance on large-scale datasets.

\textbf{Popularity Smoothing Factor \(\alpha\).}
Figure~\ref{fig:hyper_comp_big} (c) and (d) show that both ML-1M and Yelp favor relatively small smoothing factors. The best performance is achieved when \(\alpha \in [0.1, 0.4]\), while increasing \(\alpha\) to 0.8 leads to a consistent performance drop. This trend indicates that excessive smoothing weakens the discriminative power of in-community popularity signals on large-scale data.

\subsection{Fair warm-up protocol}

In some cases, a model may benefit from continued optimization even after apparent convergence during pretraining. To ensure a fair and controlled comparison between different negative sampling strategies, we adopt a unified warm-up protocol for all methods.

Specifically, we first pretrain the model using a common baseline sampler until convergence. Starting from the same pretrained checkpoint, we then perform a second-stage fine-tuning procedure, during which only the negative sampling strategy is changed while all other training configurations remain identical. This two-stage pipeline isolates the effect of negative sampling from initialization and optimization factors.

\begin{figure}
    \centering
    \includegraphics[width=0.9\linewidth]{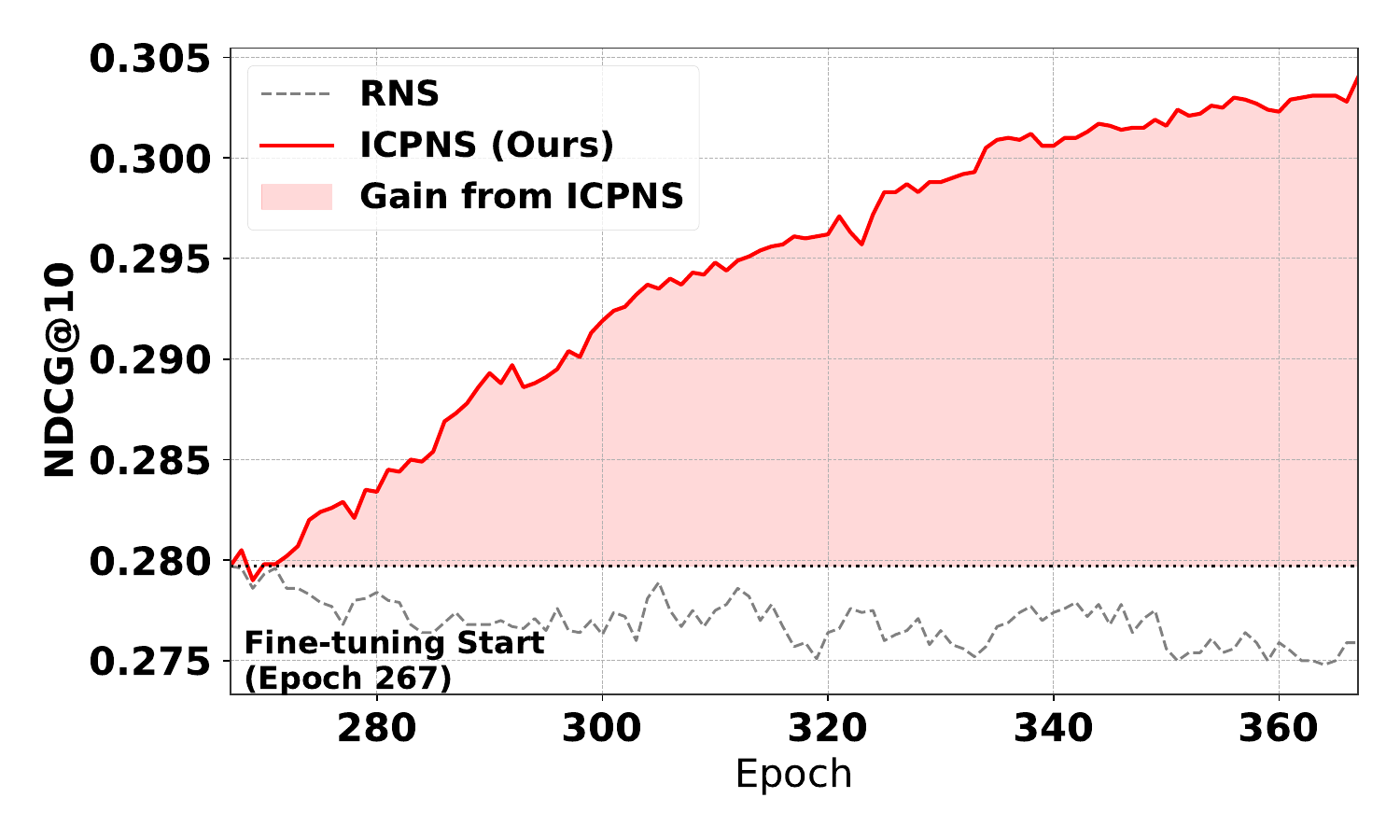}
    \caption{NDCG@10 trajectories during fine-tuning on ML-100K. Both methods start from the same pretrained checkpoint at Epoch 267. ICPNS consistently improves performance over RNS under the same warm-up protocol.}
    \label{fig:impv}
\end{figure}

Under this protocol, we compare RNS and ICPNS using LightGCN as a fixed backbone by measuring the relative improvement in NDCG@10 on the ML-100K dataset during the fine-tuning stage. As illustrated in Figure\ref{fig:impv}, both methods start from the same performance level at the beginning of fine-tuning, ensuring that any subsequent performance divergence can be attributed solely to the difference in negative sampling strategies rather than discrepancies in model warm-up or convergence behavior.

\section{Metrics}
\subsection{Recommendation Metrics}
\label{sec:metrics}

In this work, we evaluate recommendation performance using standard
top-$K$ ranking metrics, including Recall@K, NDCG@K, MRR@K, and Precision@K.
These metrics characterize complementary aspects of recommendation quality,
such as coverage of relevant items, ranking sensitivity, early-hit behavior,
and recommendation accuracy.

Let $U$ denote the set of users.
For a user $u \in U$, let $R(u)$ be the set of ground-truth relevant items,
and $\hat{R}(u)$ be the ranked recommendation list truncated at $K$.
We denote $|\cdot|$ as set cardinality.

\subsubsection{Recall@K}

Recall@K measures the fraction of relevant items that are successfully retrieved
within the top-$K$ recommendation list.
It is defined as
\begin{equation}
\mathrm{Recall@K}
=
\frac{1}{|U|}
\sum_{u \in U}
\frac{|\hat{R}(u) \cap R(u)|}{|R(u)|}.
\end{equation}
Recall focuses on coverage of relevant items and does not consider their
exact ranking positions.

\subsubsection{NDCG@K}

Normalized Discounted Cumulative Gain (NDCG) evaluates ranking quality by
accounting for the positions of relevant items, assigning higher importance to
items ranked near the top.
NDCG@K is defined as
\begin{equation}
\mathrm{NDCG@K}
=
\frac{1}{|U|}
\sum_{u \in U}
\left(
\frac{
\sum_{i=1}^{K} \mathbbm{1}(i \in R(u)) \cdot \frac{1}{\log_2(i+1)}
}{
\sum_{i=1}^{\min(|R(u)|,K)} \frac{1}{\log_2(i+1)}
}
\right),
\end{equation}
where $\mathbbm {1}(\cdot)$ is the indicator function.
The denominator corresponds to the ideal DCG obtained by ranking all relevant
items at the top positions.

\subsubsection{MRR@K}

Mean Reciprocal Rank (MRR) measures how early the first relevant item appears
in the ranked list.
It is defined as
\begin{equation}
\mathrm{MRR@K}
=
\frac{1}{|U|}
\sum_{u \in U}
\frac{1}{\mathrm{rank}^{*}_{u}},
\end{equation}
where $\mathrm{rank}^{*}_{u}$ denotes the rank position of the first relevant
item for user $u$ within the top-$K$ list.
If no relevant item appears in the top-$K$ recommendations, the contribution
for that user is set to zero.
MRR emphasizes early-hit performance.

\subsubsection{Precision@K}

Precision@K measures the fraction of recommended items that are relevant.
It evaluates the accuracy of the top-$K$ recommendation list and is defined as
\begin{equation}
\mathrm{Precision@K}
=
\frac{1}{|U|}
\sum_{u \in U}
\frac{|\hat{R}(u) \cap R(u)|}{|\hat{R}(u)|}.
\end{equation}
Precision@K focuses on the purity of the recommendation list and complements
Recall@K by reflecting how many recommended items are relevant.

\subsection{Clustering Metrics}
\label{app:clustering_metrics}

To evaluate the quality of user community structures, we additionally adopt
two widely used internal clustering validation metrics: the Silhouette score
and the Calinski--Harabasz index. Both metrics assess clustering compactness
and separation without requiring external labels.

\subsubsection{Silhouette}

The Silhouette score measures how well each sample fits within its assigned
cluster compared to other clusters.
For a data point $i$, let $a(i)$ denote the average distance between $i$ and
all other points in the same cluster, and let $b(i)$ denote the minimum average
distance between $i$ and points in any other cluster.
The Silhouette coefficient for $i$ is defined as
\begin{equation}
s(i) = \frac{b(i) - a(i)}{\max\{a(i), b(i)\}}.
\end{equation}
The overall Silhouette score is obtained by averaging $s(i)$ over all samples.

\subsubsection{Calinski--Harabasz Index}

The Calinski--Harabasz (CH) index evaluates clustering quality by comparing
between-cluster dispersion with within-cluster dispersion.
It is defined as
\begin{equation}
\mathrm{CH}
=
\frac{\mathrm{Tr}(B_k)}{\mathrm{Tr}(W_k)}
\cdot
\frac{n - k}{k - 1},
\end{equation}
where $n$ is the total number of samples, $k$ is the number of clusters,
$\mathrm{Tr}(B_k)$ denotes the trace of the between-cluster dispersion matrix,
and $\mathrm{Tr}(W_k)$ denotes the trace of the within-cluster dispersion matrix.

\newpage

\end{document}